\documentclass[11pt]{article}

\usepackage{graphicx}
\usepackage{bm}

\usepackage{enumitem}
\usepackage{enumerate}
\usepackage{multirow}
\usepackage{times}

\usepackage{caption}

\usepackage{amssymb}
\usepackage{amstext}
\usepackage{amsfonts}
\usepackage{amsmath}
\usepackage{amsthm}

\usepackage{url}

\usepackage{chemarr}
\usepackage{bbm}

\usepackage{newfloat,algcompatible}
\usepackage[size=small]{caption}
\usepackage{etoolbox}

\AtEndEnvironment{algorithm}{\noindent\hrulefill\par\nobreak\vskip-3pt}

\DeclareFloatingEnvironment[
    fileext=loa,
    listname=List of Algorithms,
    name=Algorithm,
    placement=!tbhp,
]{algorithm}
\DeclareCaptionFormat{algorithms}{\vskip-3pt\hrulefill\par#1#2#3\vskip-3pt\hrulefill}
\algblock[Procedure]{Procedure}{EndProcedure}
\algblockdefx[Procedure]{Procedure}{EndProcedure}%
    [1]{\textbf{procedure:} #1}%
    {\textbf{end procedure}}

\begin{document}

\title{Efficient Finite Difference Method for Computing Sensitivities of Biochemical Reactions}

\author{
Vo Hong Thanh\footnote{Aalto University, Department of Computer Science. Email: thanh.vo@aalto.fi}, Roberto Zunino\footnote{University of Trento, Department of Mathematics. Email: roberto.zunino@unitn.it} and Corrado Priami\footnote{The Microsoft Research -- University of Trento Centre for Computational and Systems Biology (COSBI), Italy and University of Pisa, Department of Computer Science. Email: priami@cosbi.eu} }

\maketitle
\begin{abstract}
Sensitivity analysis of biochemical reactions aims at quantifying the dependence of the reaction dynamics on the reaction rates. The computation of the parameter sensitivities, however, poses many computational challenges when taking stochastic noise into account. This paper proposes a new finite difference method for efficiently computing sensitivities of biochemical reactions. We employ propensity bounds of reactions to couple the simulation of the nominal and perturbed processes. The exactness of the simulation is preserved by applying the rejection-based mechanism. For each simulation step, the nominal and perturbed processes under our coupling strategy are synchronized and often jump together, increasing their positive correlation and hence reducing the variance of the estimator. The distinctive feature of our approach in comparison with existing coupling approaches is that it only needs to maintain a single data structure storing propensity bounds of reactions during the simulation of the nominal and perturbed processes. Our approach allows to compute sensitivities of many reaction rates simultaneously. Moreover, the data structure does not require to be updated frequently, hence improving the computational cost. This feature is especially useful when applied to large reaction networks. We benchmark our method on biological reaction models to prove its applicability and efficiency.
\end{abstract}

%
%
\section{Introduction}
\label{sec:1}
Biochemical reactions at cellular level are inherently nonlinear and stochastic due to the discreteness in the copy numbers of molecular species and randomness in molecular collisions enabling the reactions between these species. The stochasticity of biochemical reactions (often referred to as biological noise) can be amplified when key species like genes, mRNAs are often present with low copy numbers. The effects of biological noise have been demonstrated to play an important role in driving biological processes like gene regulation~\cite{MA_1,MA_2,VKBL_1,OTKGO_1,ELSS_1} or cell fate decision~\cite{ARM_1}. The noise may be further propagated across cells leading to remarkable diversity at organism level~\cite{PO_1,RO_1}. 

Stochastic chemical kinetics has been adopted to study dynamical behavior of biochemical reactions where stochastic noise is treated as an intrinsic part. It acknowledges the discrete nature of molecular species by keeping track the discrete copy number of each species, called {\em population}. The collection of populations of species forms the {\em system state}. The possibility that a reaction occurs in the next infinitesimal time is assigned a probability which is proportional to a {\em propensity} function. The propensity of a reaction depends on the population numbers of reactant species and on its reaction rate. The probability distribution of the system state over time is characterized by the chemical master equation (CME)~\cite{Gi_1} and can be exactly realized by the Gillespie's stochastic simulation algorithm (SSA)~\cite{Gi_2,Gi_3}. The core of SSA is a Monte Carlo procedure that moves the system state by randomly selecting a reaction to fire according to its propensity. Two first implementations of the Monte Carlo step are the direct method (DM) and first reaction method (FRM)~\cite{Gi_2}. Since then, many efficient implementations of the Monte Carlo step have been introduced including DM with improved search~\cite{CLP_1,MPCSS_1,TZP_2017}, with tree-based search~\cite{BBS_1,TZ_1,TZ_2}, with composition-rejection search~\cite{STP_1} and with partial-propensity approach~\cite{RSS_1}, the next reaction method (NRM)~\cite{GB_1,And_1}, the rejection-based SSA (RSSA)~\cite{TPZ_1,TZP_1,TZP_2,Thanh_2018,Thanh_1} and other improvements~\cite{Gi_4,YGP_1,ACK_1,TZP_3,Marchetti_2016,LP_2,TZ_3,MPT_2017}. Extensions of SSA have been introduced to cope with different aspects of biochemical reactions like time delays~\cite{BVTH_1,BBLT_1,Cai_1,And_1,TZP_2017} and time-dependent reaction rates~\cite{LVTH_1,And_1,TP_1,TMRP_1}.

The dynamical behavior of biochemical systems is affected by reaction rates. The changes in the rates of reactions, hence reaction propensities, due to, for example, changes in the cellular environment and/or measurement errors, may significantly alter the system behavior. Therefore it is important to quantify the dependence of the reaction dynamics on their rates. Sensitivity analysis aims at quantitatively characterizing this dependency. Different methods for sensitivity analysis of stochastic chemical kinetics have been introduced including finite differences~\cite{RSK_2010,Anderson_2012,SAR_2013,MII_2017}, likelihood ratios~\cite{PA_2007,WA_2012,MOV_2012}, infinitesimal perturbation analysis~\cite{SRK_2012} and other~\cite{GK_2014}. Each of these methods has its own advantages and drawbacks (see Asmussen and Glynn~\cite{AG_2007} for a general discussion on these methods). 

The finite difference scheme measures the difference in the behavior of reactions by imposing a small perturbation to reaction rates around their nominal values. A crude estimator for computing sensitivities based on the finite difference approach is to use two independent SSA simulation runs in which the first one simulates reactions with their nominal rate values and the second one applies to perturbed values, respectively. The simulation requires two streams of independent random numbers, resulting in large variance of the crude estimator. To reduce the variance of the finite difference estimator, Rathinam et al.~\cite{RSK_2010} introduced the common random number (CRN) method where the same stream of random numbers is used for both the simulations of reactions with nominal and perturbed rates. The correlation of the nominal and perturbed processed introduced by CRN, however, will be broken if the simulation time is large. Anderson~\cite{Anderson_2012} recently introduced the coupled finite difference (CFD) which improves the estimator by coupling the nominal and perturbed processes by exploiting the random time change representation in which the number of firings of reactions is modeled as unit-rate Poisson processes with integrated propensities. We note that Rathinam et al.~\cite{RSK_2010} also proposes a method that employs the random time change representation, but is often more involved than CFD. For each reaction, CFD splits the Poisson processes associated with the reaction in the nominal and perturbed processes so that a common Poisson process having smallest rate will be shared by these processes. CFD thus requires an additional computational cost for maintaining and sampling the shared Poisson processes. This extra computational cost is increasing with the number of reactions, which negatively affects the performance of the CFD when applying for large models. We also remark that for CRN and CFD, if the sensitivities with respect to several reaction rates are required, the computation must be performed for each rate separately. 

In this paper, we propose a new finite difference scheme, called the rejection-based finite difference (RFD) method, to address the drawbacks of the state-of-the-art coupling strategies for sensitivity analysis of biochemical reactions. RFD constructs the estimator by using propensity bounds of reactions to couple the nominal and perturbed processes and employing the rejection-based simulation approach by Thanh et al.~\cite{TPZ_1} to simulate these processes. The propensity bound of a reaction is an interval that encloses the propensity values of the reaction in both the nominal and perturbed processes. The propensity bound of each reaction is derived by bounding the population of reactant species and its rate values in these processes. Employing the propensity bounds of reactions and the rejection-based simulation allows RFD to correlate and synchronize the selection of reactions in the nominal and perturbed processes in each step. Specifically, for each simulation step, a reaction is selected as a candidate for firing in both the nominal and perturbed processes using its propensity bound. Knowing the candidate reaction, each process evaluates the exact propensity value of the candidate with its own state and decides whether to fire the candidate by applying the rejection-based test. The simulation of the nominal and perturbed processes often fire the candidate reaction and then jump their states together, increasing the positive correlation between them and hence reducing the variance of the estimator. The use of propensity bounds of reactions to implement the finite difference estimator offers many computational advantages in comparison with CRN~\cite{RSK_2010} and CFD~\cite{Anderson_2012}. First, we only need a single data structure to store propensity bounds of reactions during the simulation of the nominal and perturbed processes. This enables RFD to compute sensitivities by simultaneously perturbing many reaction rates in a run, instead of one at a time as in CRN and CFD. We remark that the exactness of the marginal distributions of the processes by RFD is preserved by the application of the rejection-based mechanism. Second, our approach does not need to simulate and maintain additional processes as in CRN and CFD, which is often computationally expensive for large models. In addition, the propensity bounds of reactions in RFD are only updated infrequently. This computational efficiency of our approach makes it especially useful for large models.  

The paper is organized as follows. Section~\ref{sec:2} reviews the background of the stochastic kinetics and sensitivity analysis of biochemical reactions using stochastic simulation. Section~\ref{sec:3} presents our new finite difference scheme that uses the concept of propensity bounds and the rejection-based approach for computing sensitivities. The general framework for stochastic simulation of biochemical reactions using the rejection-based technique is recalled in Section~\ref{sec:3.1}. The description of the rejection-based coupling strategy is in Section~\ref{sec:3.2} and its detailed implementation is outlined in Section~\ref{sec:3.3}. Section~\ref{sec:4} presents the experimental results of the application of our approach to biological reactions models considered as benchmarks. The concluding remarks are in section~\ref{sec:5}.    

\section{Stochastic chemical kinetics}
\label{sec:2}
We consider a well-mixed reactor volume consisting of $n$ molecular species represented by $S_i$ for $i=1, \ldots, n$. The exact population of each species $S_i$ at a time $t$ is kept track and denoted by $X_i(t)$. The collection of population of species at time $t$ forms the system state and is expressed by an $n$-vector $X(t) = (X_{1}(t), ..., X_n(t))$. 

Species in the reactor volume can interact with other species through $m$ reactions. A reaction $R_{j}$ for $j = 1\ldots m$ describes a possible combination of species in a unidirectional way to produce other species.
\begin{equation}
R_{j}: v_{1j}S_{1} + ... + v_{nj}S_{n} \stackrel{c_{j}}{\rightarrow} v'_{1j}S_{1} + ... + v'_{nj}S_{n}
\label{equ:1}
\end{equation}
where the species on the left side of the arrow are called reactants and the ones on the right side are called products. The non-negative integer $v_{ij}$ and $v'_{ij}$, called {\em stoichiometric coefficients}, denote the number of molecules of a reactant consumed and the number of a product produced by firing $R_j$, respectively. Each reaction is associated with a parameter $c_{j}$ which is called the (stochastic) reaction rate constant.

Each reaction $R_j$ in the stochastic chemical kinetics framework is quantified by two quantities: a {\em state change vector} $v_j$ and a {\em propensity} function $a_j$. The state change vector $v_j$ denotes the change amount in the system state $X(t)$ caused by reaction $R_j$ when it is selected to fire. The state change vector $v_j$ of a reaction $R_j$ is an $n$-vector where the $i$th element is $v'_{ij} - v_{ij}$. The reaction propensity $a_j$ quantifies the likeliness that a reaction $R_j$ occurs per unit time~\cite{Gi_2}. Specifically, the probability that a reaction $R_j$ fires in the next infinitesimal time $t + dt$ is $a_j(X(t))dt$, given the current state $X(t)$ at time $t$. An exact form of the propensity function $a_j$ is dependent on reaction kinetics applied. For standard mass-action kinetics, the propensity $a_j$ of reaction $R_j$ is proportional its reactants and reaction rate constant $c_j$ and can be computed as:
\begin{equation}
   a_j(X(t)) = c_j h_j(X(t))
\label{eq:propensity}
\end{equation}
where $h_j(X(t))$ counts the number of distinct combinations of reactants involved in $R_j$. The number of combinations of reactants of {\em synthesis reactions}, where species are produced from an external reservoir, is defined $h_j(X(t)) = 1$. 

The dynamics of state $X(t)$ under the stochastic chemical kinetics framework is modeled as a (continuous-time) jump Markov process and can be fully described by the chemical master equation (CME)~\cite{Gi_1}. The exact stochastic simulation algorithm (SSA)~\cite{Gi_2,Gi_3} can be applied to realize $X(t)$. SSA is an exact algorithm in the sense that it does not introduce approximation in the sampling. The mathematical background for the simulation of SSA is the joint probability density function (pdf) $p(\tau, \mu)$ such that $p(\tau, \mu)d\tau$ gives the probability that a reaction $R_{\mu}$ fires in the next infinitesimal time $t + \tau + d\tau$, given the state $X(t)$ at time $t$. Its closed form is: 
\begin{equation}
  p(\tau,\mu) = a_{\mu}exp(-a_{0}\tau)
\label{equ:pdf}
\end{equation}
where $a_{0} = \sum_{j=1}^{m}a_{j}$. 

Various Monte Carlo strategies have been introduced for SSA in order to sample the pdf $p(\tau,\mu)$~\cite{Gi_2,GB_1,TPZ_1}. One approach is the direct method (DM) which samples the pdf $p(\tau,\mu)$ in Eq.~\ref{equ:pdf} by using the fact that reaction $R_{\mu}$ fires with a discrete probability $a_\mu/a_0$ and the firing time $\tau$ is exponentially distributed with rate $a_0$. Thus, for each simulation iteration, $m$ propensities $a_j$ for $j = 1\ldots m$ and their sum $a_0 = \sum_{j=1}^{m}a_j$ are computed. The next reaction firing $R_{\mu}$ with probability $a_\mu/a_0$ is selected by
\begin{equation}
  \mu = \textrm{smallest reaction index such that: }\sum^{\mu}_{j=1}a_{j} \geq r_{1}a_{0} 
\label{equ:mu-dm}
\end{equation}
and the firing time $\tau$ is generated as 
\begin{equation}
  \tau = \frac{1}{a_{0}}\ln \left( \frac{1}{r_{2}} \right)
\label{equ:tau-dm}
\end{equation}
where $r_{1}$ and $r_{2}$ are two random numbers generated from a uniform distribution $U(0,1)$. The state is moved to a new state $X(t + \tau) = X(t) + v_\mu$. Propensities are updated as well to reflect the change in the system state. In practice, a reaction dependency graph~\cite{GB_1} is often employed to reduce the number of reaction propensities. The simulation is repeated to form a simulation trajectory until a specified ending time is reached. 

An alternative approach to construct realizations of $X(t)$ is to use the next reaction method (NRM) by employing the random time change (RTC) representation~\cite{And_1}. Let $Po_j$ be independent unit-rate Poisson processes denoting the number firings of reactions $R_j$, with $j = 1, \ldots, m$. The dynamics of $X(t)$ under RTC can be expressed as 
\begin{equation}
   X(t) = X(0) + \sum_{j=1}^{m} Po_j\Big(\int_{0}^{t}a_j(X(s))ds\Big)v_j
\label{eq:rtc-representation}
\end{equation}
Using the RTC representation, each reaction $R_j$ has an internal time $T_j$, which determines the starting time of the clock until reaction $R_j$ fires at the next time $t + \tau_j$, given the current time $t$. Let $P_j$ be the next firing time of reaction $R_j$ assuming that $a_j$ is not changing, Then, the waiting time until its firing is $\tau_j = (1/a_j)(P_j - T_j)$. Because $Po_j$ is a unit Poisson process, the next firing time $P_j$ is generated by sampling an exponential distribution with rate $1$. Having the waiting times of all reactions, the reaction $R_\mu$ having smallest waiting time $\tau = \min_{j=1}^{m}(\tau_j)$ is selected to fire. By explicitly keeping track of the firing times of Poisson processes, NRM consumes only one random number per simulation step.

\subsection{Sensitivity analysis of stochastic chemical kinetics}
This section deals with sensitivity analysis using stochastic simulation. We recall two finite difference schemes proposed recently for computing sensitivities: the common random number (CRN)~\cite{RSK_2010} and coupled finite difference (CFD)~\cite{Anderson_2012}. These methods derive the sensitivity of the system dynamics by applying a small perturbation to a reaction rate constant. The computation of sensitivities with respect to several reaction rates requires the computation being performed for each rate separately. 

Let $c$ be an $m$-vector in which the $j$th element is the reaction rate constant $c_j$ of a reaction $R_j$ for $j = 1, \ldots m$. We denote the system state at time $t$ corresponding to rate vector $c$ with $X^{c}(t)$. Let $f$ be a function of the state which represents a measurement of interest. The quantity $S(c)$ that we want to measure is defined as: 
\begin{equation}
   S(c) = \mathbb{E}[f(X^{c}(t))]
\end{equation}
where $\mathbb{E}[-]$ denotes the expectation operator.

Let $R_k$ be the reaction for which we want to quantify the dependence of $S(c)$ on its reaction rate constant $c_k$. The aim of sensitivity analysis is to compute the partial derivative (called {\em sensitivity coefficient}) of $S(c)$ with respect to the reaction rate $c_k$, i.e.,~${\partial S(c)}/{\partial c_k}$. The sensitivity coefficient ${\partial S(c)}/{\partial c_k}$ can be estimated by applying a small scalar perturbation $\epsilon_k$ to the reaction rate $c_k$. Let $e_k$ be a unit $m$-vector in which the $k$th element is $1$, while other elements are $0$s. The sensitivity coefficient with respect to a reaction rate $c_k$ can be approximated by the {\em forward difference}
\begin{equation}
    \frac{\partial S(c)}{\partial c_k} \approx \frac{S(c + \epsilon_k e_k) - S(c)}{\epsilon_k}		                                   \approx \frac{\mathbb{E}[f(X^{c+\epsilon_ke_k}(t))] - \mathbb{E}[f(X^{c}(t))]} {\epsilon_k}
\label{parameter-sensitivity-problem}
\end{equation}
The bias of the forward difference due to the truncation error is $O(\epsilon_k)$. We note that the bias can be reduced to $O(\epsilon^2_k)$ by using the {\em centered finite difference} method~\cite{AG_2007}. In Eq.~\ref{parameter-sensitivity-problem}, the bias becomes zero in the limit that $\epsilon_k \rightarrow 0$.

The estimator for the forward difference in Eq.~\ref{parameter-sensitivity-problem} can be constructed as
\begin{equation}
  Z = \frac{1}{N}\sum_{i=1}^{N}\frac{f(X_{[i]}^{c+\epsilon_ke_k}(t)) - f(X_{[i]}^{c}(t))}{\epsilon_k} 
\label{sensitivity-estimator}
\end{equation}
where $N$ is the number of simulation runs and $X_{[i]}^{c}(t)$ denotes the $i$th realization with rate parameter $c$. A naive implementation of the estimator where $X_{[i]}^{c}(t)$ and $X_{[i]}^{c+\epsilon_ke_k}(t)$ are generated independently will produce a large variance. In fact, the variance of the estimator $var[Z]$ in the naive implementation is equal the sum of two variances $var[f(X^{c+\epsilon_ke_k}(t))]$ and $var[f(X^{c}(t))]$ because their covariance is zero (i.e.,~$cov[f(X^{c+\epsilon_ke_k}(t)), f(X^{c}(t))] = 0$). CRN and CFD reduce the variance of the estimator by introducing a (positive) correlation between the $X^{c}(t)$ and $X^{c+\epsilon_ke_k}(t)$, hence $cov[f(X^{c+\epsilon_ke_k}(t)), f(X^{c}(t))] \neq 0$, during the simulation of these processes. 

The CRN method correlates $X^{c}(t)$ and $X^{c+\epsilon_ke_k}(t)$ by using the same stream of random numbers during the simulation. Algorithm~\ref{alg:crn} outlines the steps of the CRN approach as applied to SSA. The key of the CRN is that the random number generator used in both the simulations of $X^{c}(t)$ and $X^{c+\epsilon_ke_k}(t)$ is initialized with the same seed w.

\begin{algorithm}
\caption{Common Random Number method (CRN)} \label{alg:crn}
\begin{algorithmic}[1]
\STATE define reaction index $k$ for sensitivity analysis
\STATE initialize time $t = 0$ and state $X^{c} = X^{c+\epsilon_ke_k} = x_0$
\STATE generate a random seed w
\STATE seed the random number generator with w
\STATE realize $X^{c}$ by performing SSA until time $T_{max}$
\STATE reseed the random number generator with w
\STATE realize $X^{c+\epsilon_ke_k}$ by performing SSA until time $T_{max}$
\end{algorithmic}
\end{algorithm}
  
The CFD employs the RTC representation in Eq.~\ref{eq:rtc-representation} to correlate the nominal process $X^{c}(t)$ and its perturbed one $X^{c+\epsilon_ke_k}(t)$. Specifically, by RTC representation and additive property of the Poisson process, it gives
\begin{equation}
X^{c}(t) = X(0) + \sum_{j=1}^{m}Po_{j,1}\Big(\int_{0}^{t}b_j(s)ds \Big)v_j 
         + \sum_{j=1}^{m}Po_{j,2}\Big(\int_{0}^{t}(a_j(X^{c}(s)) - b_j(s))ds \Big)v_j
\label{RTC-nominal-process}
\end{equation}
and 
\begin{equation}
X^{c + \epsilon_ke_k}(t) = X(0) + \sum_{j=1}^{m}Po_{j,1}\Big(\int_{0}^{t}b_j(s)ds \Big)v_j 
										     + \sum_{j=1}^{m}Po_{j,3}\Big(\int_{0}^{t}(a_j(X^{c + \epsilon_ke_k}(s)) - b_j(s))ds\Big)v_j
\label{RTC-perturbed-process}
\end{equation}
where $b_j(t) = \min(a_j(X^{c}(t)), a_j(X^{c + \epsilon_ke_k}(t)))$ and $Po_{j,i}$ for $j = 1,\ldots, m$ and $i \in \{1, 2, 3\}$ denote independent unit-rate Poisson processes. The representation of the nominal and perturbed processes in Eqs.~\ref{RTC-nominal-process} -~\ref{RTC-perturbed-process} is the key of the CFD method outlined in Algorithm~\ref{alg:cfd} where the simulation of the Poisson processes $Po_{j,1}(\int_{0}^{t}b_j(s)ds)$ is shared. 

\begin{algorithm}
\caption{Coupled Finite Difference method (CFD)} \label{alg:cfd}
\begin{algorithmic}[1]
\STATE define reaction index $k$ for sensitivity analysis
\STATE initialize time $t = 0$ and state $X^{c} = X^{c+\epsilon_ke_k} = x_0$
\STATE set $T_{j,i} = 0$ and $P_{j,i} = \ln(1/r_{j,i})$ where $r_{j,i} \sim U(0,1)$ for $j = 1,\ldots, m$ and $i \in \{1, 2, 3\}$
\WHILE{($t < T_{max}$)}     
	\STATE compute $a_j(X^{c+\epsilon_ke_k})$ and $a_j(X^{c}(t))$ for $j = 1,\ldots, m$
	\STATE set $b_{j,1} =\min(a_j(X^{c+\epsilon_ke_k}), a_j(X^{c}(t)))$ and compute $b_{j,2} = a_j(X^{c}) - b_{j,1}$ and $b_{j,3} = a_j(X^{c+\epsilon_ke_k}) - b_{j,1}$ for $j = 1,\ldots, m$
	\STATE compute $\tau_{j,i} = (P_{j,i} - T_{j,i}) / b_{j,i}$ for $j = 1,\ldots, m$ and $i = 1,\ldots, 3$
	\STATE set $\tau = \min(\tau_{j,i})$ and let $(\mu,\alpha)$ be the pair of indices where the minimum is selected
	\STATE set $t = t + \tau$
	\IF {$(\alpha == 1)$}
			\STATE update $(X^{c}, X^{c+\epsilon_ke_k}) = (X^{c}, X^{c+\epsilon_ke_k}) + (v_\mu, v_\mu)$	
	\ELSIF{$(\alpha == 2)$}
			\STATE update $X^{c} = X^{c} + v_\mu$
	\ELSIF{$(\alpha == 3)$}
			\STATE update $X^{c+\epsilon_ke_k} = X^{c+\epsilon_ke_k} + v_\mu$
	\ENDIF				
	\STATE set $T_{j,i} = T_{j,i} + b_{j,i}\tau$ for $j = 1, \ldots, m$ and $i = \{1, 2, 3\}$
	\STATE set $P_{\mu,\alpha} = P_{\mu,\alpha} + \ln(1/r)$ where $r \sim U(0,1)$
\ENDWHILE

\end{algorithmic}
\end{algorithm}

\section{Sensitivity analysis using rejection-based approach}
\label{sec:3}
This section introduces a new finite difference scheme that employs the concept of propensity bounds and rejection-based simulation technique for efficiently computing sensitivities of biochemical reactions. We first recall the principle of the rejection-based stochastic simulation algorithm (RSSA), which provides the theoretical framework for our new coupling strategy. Then, we describe in detail how our new rejection-based finite difference (RFD) method correlates the simulation of the nominal and perturbed processes to construct the estimator. We also compare our coupling strategy with the CRN and CFD methods discussed in previous section to highlight the advantages of our approach. 

\subsection{Background on rejection-based simulation}
\label{sec:3.1}
RSSA, originally proposed by Thanh \textit{et al.}~\cite{TPZ_1}, is an exact stochastic simulation algorithm that aims to improve simulation performance by reducing the average number of propensity calculations during the simulation. The exactness proof of RSSA as well as its extensions for handling complex reaction mechanisms can be accessed through the recent work~\cite{TZP_2017,TPZ_1,Marchetti_2016,TP_1,TMRP_1}. The mathematical framework for the selection of reaction firings in RSSA is the rejection-based sampling technique. It uses propensity bounds $[\underline{a_j}, \overline{a_j}]$, which delimits all possible values of the propensity $a_j(X(t))$ of each reaction $R_j$ with $j = 1,\ldots, m$, to select the next reaction firing. The propensity bounds are computed by bounding the state $X(t)$ to the fluctuation interval $[\underline{X}, \overline{X}]$ such that the inequality $\underline{X} \leq X(t) \leq \overline{X}$ holds for each species $S_i$, with $i = 1,\ldots n$, in the state $X(t)$. 

RSSA selects the next reaction firing using propensity bounds in two steps. First, a candidate reaction $R_\mu$ is selected with probability $\overline{a_\mu}/\overline{a_0}$ where $\overline{a_0} = \sum_{j = 1}^{m}\overline{a_j}$. Second, the candidate $R_\mu$ is validated through a rejection test with success probability $a_\mu(X(t)) / \overline{a_\mu}$. If the candidate $R_\mu$ passes the rejection test, then it is accepted to fire. Otherwise, it is rejected and another candidate is selected to test. The rejection test requires to compute propensity $a_\mu(X(t))$, but the implementation can postpone the computation by exploiting the fact that if a candidate reaction $R_\mu$ is accepted with probability $\underline{a_\mu}/\overline{a_\mu}$, then it can be accepted without evaluating $a_\mu(X(t))$ because of $\underline{a_\mu}/\overline{a_\mu} < a_\mu(X(t))/\overline{a_\mu}$.

RSSA also uses the propensity bounds of reactions to compute the firing time of the accepted candidate reaction. Let $l$ be the number of trials such that the first $l - 1$ trials are rejections and at the $l$th trial $R_\mu$ is accepted. The firing time $\tau$ of the accepted candidate $R_\mu$ in RSSA is the sum of $l$ independent exponentially distributed numbers with the same rate $\overline{a}_0$, which is equivalent to an $\texttt{Erlang}(l, \overline{a_0})$ distribution. RSSA thus generates the firing time $\tau$ of the accepted candidate $R_\mu$ by sampling the corresponding $Erlang$ distribution. 

\subsection{Coupling by rejection-based approach}
\label{sec:3.2}
We describe in this section the principle of the rejection-based finite difference (RFD) method that employs the rejection-based framework introduced by RSSA for coupling the simulation of the original and perturbed processes, with the aim of computing sensitivities. Let $k$ be the index of reaction $R_k$, for which we  want to measure the sensitivity. Let $X^{c}(t)$ be the state of the nominal process and $X^{c + \epsilon_ke_k}(t)$ be state of the perturbed process obtained by changing $k$th element of the rate vector $c$ by an amount denoted by $\epsilon_ke_k$. Consider a time interval $[0, t]$. Let $\overline{a}_j$ be an arbitrary propensity upper bound that is greater than the exact propensity values $a_j(X^c(t))$ and $a_j(X^{c+\epsilon_ke_k}(t))$ of each reaction $R_j$, with $j = 1, \ldots, m$, during the time interval, i.e., $\overline{a}_j \geq a_j(X^c(s))$ and $a_j(X^{c+\epsilon_ke_k}(s))$ for all $s \in [0, t]$. Let $Po_j$ be a unit-rate Poisson process, evaluated over the time interval $[0, t]$. By the addictive property of the Poisson distribution, the Poisson process with rate $\overline{a}_j$ can be decomposed as
\begin{equation}
  Po_j\Big(\overline{a}_j t\Big) = Po_{j,1}\Big(\int_{0}^{t}a_j(X^c(s))ds\Big) + Po_{j,2}\Big(\overline{a}_j t - \int_{0}^{t}a_j(X^c(s))ds\Big)
\end{equation}
and
\begin{equation}
  Po_j\Big(\overline{a}_j t\Big) = Po_{j,3}\Big(\int_{0}^{t}a_j(X^{c+\epsilon_ke_k}(s))ds\Big) + Po_{j,4}\Big(\overline{a}_j t - \int_{0}^{t}a_j(X^{c+\epsilon_ke_k}(s))ds\Big)
\end{equation}
Equivalently, these equations can be rewritten as
\begin{equation}
  Po_{j,1}\Big(\int_{0}^{t}a_j(X^c(s))ds\Big) = Po_j\Big(\overline{a}_j t\Big) - Po_{j,2}\Big(\overline{a}_j t - \int_{0}^{t}a_j(X^c(s))ds\Big)
\end{equation}
and
\begin{equation}
  Po_{j,3}\Big(\int_{0}^{t}a_j(X^{c+\epsilon_ke_k}(s))ds\Big) = Po_j\Big(\overline{a}_j t\Big) - Po_{j,4}\Big(\overline{a}_j t - \int_{0}^{t}a_j(X^{c+\epsilon_ke_k}(s))ds\Big)
\end{equation}

Thus, by the RTC representation, the resulting dynamics of the nominal process is 
\begin{equation}
X^{c}(t) = X(0) + \sum_{j=1}^{m}\Big[Po_j\Big(\overline{a}_j t\Big) - Po_{j,2}\Big(\overline{a}_j t - \int_{0}^{t}a_j(X^c(s))ds\Big)\Big]v_j
\label{rejection-based-nominal-process}
\end{equation}
and its perturbed one is 
\begin{equation}
X^{c + \epsilon_ke_k}(t) = X(0) + \sum_{j=1}^{m}\Big[Po_j\Big(\overline{a}_j t\Big) - Po_{j,4}(\overline{a}_j t - \int_{0}^{t}a_j(X^{c+\epsilon_ke_k})(s)ds\Big)\Big]v_j
\label{rejection-based-perturbed-process}
\end{equation}
We remark that the equality in these derivations is used in the sense of having the same distribution.

Eqs.~\ref{rejection-based-nominal-process} and~\ref{rejection-based-perturbed-process} give the mathematical basis for our coupling strategy. These equations show that if we simulate the nominal process and the perturbed process using the propensity bounds while filtering out selected candidate reactions using the corresponding exact propensities, then we can positively couple the simulation of these processes. This is made possible by the rejection-based simulation technique. We note that the original rejection-based selection in RSSA, however, does not directly apply, because there each trajectory is generated using independent random numbers, causing no correlation between them.  To make a concrete example, when using RSSA it is possible that the nominal process accepts a reaction firing after two trials, while the perturbed process only accepts the firing after four trials, whose time points are even different with respect to those for the nominal process. Instead, for computing sensitivities we need to improve the correlation, hence we need to synchronize the nominal and perturbed processes in each trial. To do this, the key idea underlying RFD is to decompose the complex rejection-based selection of RSSA into single trials and assign a time stamp for each trial. In particular, for each simulation step, a candidate reaction $R_j$ for both the processes is selected with a probability proportional to its propensity upper bound $\overline{a}_j$ and its waiting time is generated following an exponential distribution with rate $\overline{a}_0$. The rejection-based test then decides whether the candidate reaction will be accepted to fire in the corresponding process with the exact propensity $a_j(X^c(t))$ and $a_j(X^{c+\epsilon_ke_k}(t))$, respectively. Furthermore, by choosing appropriate propensity bounds for the simulation, we can make the candidate reaction to be accepted to fire in both processes frequently during the simulation, moving the states together, hence positively correlating these processes. The derivation of propensity bounds as well as the implementation of RFD are discussed in the next section. In the following, we analyze the efficiency of our rejection-based coupling strategy and compare it with CRN and CFD.  

The application of the rejection-based coupling strategy represented in Eqs.~\ref{rejection-based-nominal-process} and~\ref{rejection-based-perturbed-process} can reduce the variance of the estimator. Specifically, by subtracting $X^{c + \epsilon_ke_k}(t)$ in Eq.~\ref{rejection-based-nominal-process} and $X^{c + \epsilon_ke_k}(t)$ in Eq.~\ref{rejection-based-perturbed-process} then taking variance of the result, it shows that the variance of the estimator is proportional to the variance of Poisson processes having rate equal to the difference of the propensity upper bound and the exact propensity value. The propensity upper bounds of reactions, however, can be chosen with flexibility in which it provides us possibility for tuning RFD to improve both the variance of the estimator and the simulation performance. In particular, if we choose $\overline{a}_j = \max(a_j(X^{c}(s)), a_j(X^{c+\epsilon_ke_k}(s)))$, then our RFD approach produces an estimator that has the same variance as CFD represented in Eqs.~\ref{RTC-nominal-process} and \ref{RTC-perturbed-process}, which use the minimum value of $a_j(X^{c}(s))$ and $a_j(X^{c+\epsilon_ke_k}(s))$ to couple these processes. However, by defining the propensity bounds in this way, we have to update these bounds after each simulation step in order to maintain the constraint, hence decreasing the simulation performance. The implementation of RFD relaxes this constraint by deriving the propensity upper bounds $\overline{a}_j$ such that they can be used for many steps without the need to update, thus improving the simulation performance. We remark that the simulation results are ensured to be exact, regardless of the choice of the propensity bounds due to the rejection-based mechanism.

Although our rejection-based coupling in Eqs.~\ref{rejection-based-nominal-process} and~\ref{rejection-based-perturbed-process} involves the additional Poisson processes with rates equal to the differences of the propensity upper bound and the exact propensity values (i.e,~$Po_{j,2}(\overline{a}_j t - \int_{0}^{t}a_j(X^c(s))ds)$ and $Po_{j,4}(\overline{a}_j t - \int_{0}^{t}a_j(X^{c+\epsilon_ke_k}(s))ds)$), we do not simulate all these processes explicitly during the simulation. In fact, these residual processes are only evaluated when some candidate is accepted. This is the distinctive feature of our RFD approach, compared with the CRN and CFD coupling strategies, which either have to simulate both the nominal and perturbed processes separately, as in CRN, or require to explicitly simulate and maintain additional Poisson processes, as in CFD. We remark that the additional computational cost introduced by the  CRN and CFD methods increases with the number of reactions in the network, which negatively affects the computational performance of these approaches when simulating large models. Our rejection-based coupling only requires to keep track of propensity bounds of reactions, which are also updated infrequently. Furthermore, the use of propensity bounds for coupling allows to simulate many distinct perturbations on rates in a single run. In this case, we only need to define a propensity upper bound that constrains the propensities of reactions for all these perturbations. Having the propensity bounds, the same rejection-based coupling strategy can be applied to couple and simulate these processes.   

\subsection{RFD algorithm}
\label{sec:3.3}
We outline in Algorithm~\ref{alg:fd-rssa} the detailed implementation of RFD for realizing the nominal and perturbed processes used in sensitivity analysis. It takes a set $\mathcal{Q}$ containing reaction indices for which sensitivities need to be computed. The algorithm will generate realizations of $X^{c}$ and $X^{c + \epsilon_ke_k}$ for reaction $k \in \mathcal{Q}$ in which $\epsilon_ke_k$ denotes the amount of change in the rate of the reaction index $k$. The simulation starts with an initial state $x_0$ at time $t = 0$ and ends at time $T_{max}$.

\begin{algorithm}
\caption{Rejection-based Finite difference method (RFD)} \label{alg:fd-rssa}
\begin{algorithmic}[1]

\STATE define set $\mathcal{Q}$ containing reaction indices for sensitivity analysis
\STATE build the species-reaction (SR) dependency graph $\mathcal{G}$ \label{sr-graph}
\STATE initialize time $t = 0$ and state $X^{c} = X^{c + \epsilon_ke_k} = x_0$ for all $k \in \mathcal{Q}$ \label{begin-init}

\STATE compute rate bounds $\underline{c_j}$ and $\overline{c_j}$ for reaction $R_j$ with $j = 1,\ldots,m$
\STATE compute fluctuation interval $[\underline{X}_i, \overline{X}_i]$ that bounds the population of species $S_i$, $i = 1\ldots n$, in all states.
\STATE compute propensity bounds $\underline{a_j}$ and $\overline{a_j}$ for reaction $R_j$ with $j = 1,\ldots,m$
\STATE set $\overline{a_0} = \sum_{j=1}^{m}\overline{a_j}$ \label{end-init}
\WHILE{$(t < T_{max})$}
 \STATE set $\textrm{UpdateSpeciesSet} = \emptyset$ \label{update-species-set}
 \WHILE{(populations of each species $S_i$ in states are in $[\underline{X}_i, \overline{X}_i]$)} \label{begin-selection-loop}	
	\STATE generate three random numbers $r_1, r_2$ and $r_3 \sim U(0,1)$
	\STATE set $\tau = (1/r_1)\ln(1/\overline{a_0})$
	\STATE update time $t = t + \tau$
	\STATE select minimum index $\mu$ s.t. $\sum^{\mu}_{j=1}\overline{a_{j}} > r_{2} \overline{a_{0}}$ 
	\IF{($r_3 \leq \underline{a_\mu} / \overline{a_\mu}$)}
		\STATE update $X^{c}$ and $X^{c + \epsilon_ke_k}$ for all $k \in \mathcal{Q}$ by $v_\mu$
	\ELSE
	  \STATE compute $a_\mu(X^{c})$
		\IF{($r_3 \leq a_\mu(X^{c}) / \overline{a_\mu}$)}
			 \STATE update $X^{c} = X^{c} + v_\mu$
		\ENDIF
		\FORALL{($k \in \mathcal{Q}$)}
		  \STATE compute $a_\mu(X^{c + \epsilon_ke_k})$
			\IF{($r_3 \leq a_\mu(X^{c + \epsilon_ke_k}) / \overline{a_\mu}$)}
				\STATE update $X^{c + \epsilon_ke_k} = X^{c + \epsilon_ke_k} + v_\mu$
			\ENDIF
		\ENDFOR	
	\ENDIF	
 \ENDWHILE \label{end-selection-loop}
 \FORALL {(species $S_i$ where population $X_i^{c}$ or $X_i^{c + \epsilon_ke_k}$, with $k \in \mathcal{Q}$, $\notin [\underline{X_i}, \overline{X_i}]$)} \label{begin-update} 
   \STATE set $UpdateSpeciesSet = UpdateSpeciesSet \cup \{S_i\}$
 \ENDFOR	
 \FORALL {(species $S_i \in \textrm{UpdateSpeciesSet}$)}   
   \STATE define a new fluctuation interval $[\underline{X_i}, \overline{X_i}]$   
   \STATE extract reactions $\textrm{ReactionsAffectedBy}(S_i)$ affected by $S_i$ from SR graph $\mathcal{G}$
	 \FORALL {($R_j \in \textrm{ReactionsAffectedBy}(S_i)$) }
		  \STATE compute new propensity bounds $\overline{a_j}$ and $\underline{a_j}$
		  \STATE update $\overline{a_0}$
   \ENDFOR
 \ENDFOR \label{end-update}	 
\ENDWHILE

\end{algorithmic}
\end{algorithm}

For the initialization in lines~\ref{begin-init} -~\ref{end-init}, RFD computes  propensity bounds $\underline{a_j}$ and $\overline{a_j}$ for each reaction $R_j$, $j = 1, \ldots, m$, such that $\underline{a_j} \leq a_j(X^{c}(t)), \{a_j(X^{c+\epsilon_ke_k}(t))\}_{k\in \mathcal{Q}} \leq \overline{a_j}$. The propensity upper bounds $\overline{a_j}$ will be used to couple the simulation of the nominal and the perturbed processes, while the propensity bounds are used to quickly accept the candidate reaction, hence avoiding computing the exact propensity value and improving the simulation performance.  

RFD computes the propensity bound $[\underline{a_j}, \overline{a_j}]$ for each reaction $R_j$, $j = 1, \ldots, m$, by bounding the rates of the reaction $R_j$ in both the nominal and perturbed processes as well as constraining all populations of reactant species in $X^{c}(t)$ and $X^{c + \epsilon_ke_k}(t)$ with $k \in \mathcal{Q}$ into the fluctuation interval $[\underline{X}, \overline{X}]$.   

For each reaction $R_j$, RFD defines a lower value $\underline{c_j}$ and an upper value $\overline{c_j}$ as the minimum and maximum of $c_j$ and $c_j + \epsilon_j$ to bound the reaction rate. Specifically, it sets $\underline{c_j} = \min(c_j, c_j + \epsilon_j)$ and $\overline{c_j} = \max(c_j, c_j + \epsilon_j)$ with $j = 1, \ldots, m$. We note that if $j \notin \mathcal{Q}$ then $\underline{c_j} = \overline{c_j} = c_j$. 

The derivation of the lower bound $\underline{X}_i$ and the upper bound $\overline{X}_i$ of populations of each species $S_i$, $i = 1,\ldots, n$, in the nominal state $X^{c}(t)$ and perturbed states $X^{c + \epsilon_ke_k}(t)$ for $k \in \mathcal{Q}$, hence forming the fluctuation interval $[\underline{X}, \overline{X}]$, is obtained by constraining the minimum and maximum population of this species in all these states. Precisely, RFD defines $X_i^{min} = \min(X_i^{c}(t), \{X_i^{c+\epsilon_ke_k}(t)\}_{k\in \mathcal{Q}})$ and $X_i^{max} = \max(X_i^{c}(t), \{X_i^{c+\epsilon_ke_k}(t)\}_{k\in \mathcal{Q}})$ as the minimum and maximum population of species $S_i$, respectively. The computation of the population bounds for $S_i$ is thus $\underline{X_i} = (1 - \delta_i)X_i^{min}$ and $\overline{X_i} = (1 + \delta_i)X_i^{max}$ where the {\em fluctuation rate} $0 \leq \delta_i \leq 1$ is a parameter. The choice of the fluctuation rate $\delta_i$ is the trade-off between reducing the update costs and increasing rejections, both affecting the simulation performance. If $\delta_i$ is too big ($\delta_i \approx 1$), we do not need to update propensity bounds often; however, it also increases the number of rejections, negatively affecting simulation performance. On the other hand, if $\delta_i$ is too small ($\delta_i \approx 0$), then we need to recompute the propensity bounds frequently, making the simulation inefficient. For typical models, the fluctuation rate chosen around $10\%$ to $20\%$ gives a good simulation performance (see numerical examples in Section~\ref{sec:5}). We note that the choice of the fluctuation rate $\delta_i$, however, does not affect the simulation result, which is always exact due to the rejection-based mechanism..

The propensity lower bound $\underline{a_j}$ and upper bound $\overline{a_j}$ for each reaction $R_j$, with $j = 1,\ldots, m$, is computed by optimizing the propensity function $a_j$ over the rate bound $[\underline{c_j}, \overline{c_j}]$ and fluctuation interval $[\underline{X}, \overline{X}]$. For mass-action propensity function $a_j$ given in Eq.~\ref{eq:propensity}, the bounds can be computed easily using its monotonic property and interval arithmetic~\cite{EKC_1}. Specifically, let $\underline{h_j}$ and $\overline{h_j}$ be the minimum and maximum of function $h_j$ over the fluctuation interval $[\underline{X}, \overline{X}]$, respectively. By monotonic property of function $h_j$, it gives $\underline{h_j} = h_j(\underline{X})$ and $\overline{h_j} = h_j(\overline{X})$. Then by interval analysis, the propensity bounds of $R_j$ can be computed as
\begin{equation}
	\underline{a_j} = \underline{c_j} \underline{h_j}
\end{equation}
and
\begin{equation}
	\overline{a_j} = \overline{c_j} \overline{h_j}
\end{equation}

The main simulation loop of the RFD algorithm using propensity bounds $[\underline{a_j}, \overline{a_j}]$ of reactions is composed of two main tasks. The first task selects the reaction to update the states using propensity bounds and the rejection-based mechanism (lines~\ref{begin-selection-loop} -~\ref{end-selection-loop}). The second task updates propensity bounds of reactions when there exists a species whose population moves out of the current fluctuation interval (lines~\ref{begin-update} -~\ref{end-update}). In addition, to facilitate the update of propensity bounds when a species whose population moves out of the fluctuation interval, the algorithm makes use of the Species-Reaction (SR) graph~\cite{TPZ_1} to retrieve which reactions should update their propensity bounds when a species exits its fluctuation interval. The SR dependency graph $\mathcal{G}$ is a directed bipartite graph which shows the dependency of reactions on species. A directed edge from a species $S_i$ to a reaction $R_j$ is in the graph if a change in the population of species $S_i$ requires reaction $R_j$ to recompute its propensity. The SR dependency graph $\mathcal{G}$ is built in line~\ref{sr-graph}.  

For the selection step in lines~\ref{begin-selection-loop} -~\ref{end-selection-loop}, a reaction firing is selected to update the states $X^{c}$ or $X^{c + \epsilon_ke_k}$ for $k \in \mathcal{Q}$ by the rejection-based selection and taking three random numbers $r_1, r_2$ and $r_3 \sim U(0,1)$ in which $r_1$ are used to compute time, while $r_2$ and $r_3$ is used to select the candidate and to validate it through a rejection test as follows. Let $\overline{a_0} = \sum_{j=1}^{m}\overline{a_j}$ be the sum of propensity upper bounds. The waiting time $\tau$ of in single trial of the rejection-based selection is an exponential distribution with rate $\overline{a_0}$. It thus can be computed by the inverse transformation as $\tau = (1/\overline{a_0})\ln(1/r_1)$. The selection of reaction firing is composed of two steps. First, candidate reaction $R_\mu$ is randomly selected with a discrete probability $\overline{a_\mu}/\overline{a_0}$. The realization of the candidate reaction $R_\mu$ can be performed by linearly accumulating propensity upper bounds until it finds the smallest reaction index $\mu$ satisfying the inequality: $\sum_{j=1}^{\mu}\overline{a_j} > r_2\cdot\overline{a_0}$ where $r_2 \sim U(0,1)$. Knowing the candidate reaction $R_\mu$, RFD applies the rejection-based test to decide whether to update the states depending on the value of the propensity of the reaction in the corresponding process. This rejection-based test ensures that marginal distributions of the nominal and perturbed processes are correct, although their joint distribution is correlated. For this purpose, RFD first checks whether $r_3 \leq \underline{a_\mu} / \overline{a_\mu}$ holds. If in fact this is the case, then both the states $X^{c}(t)$ and $X^{c + \epsilon_ke_k}$, for $k \in \mathcal{Q}$, are updated. If this test fails, RFD computes the propensities of $R_\mu$ corresponding to the each state and performs the check again with $r_3$. More in details, RFD computes the propensity $a_\mu(X^{c}(t))$ as well as $a_\mu(X^{c + \epsilon_ke_k}(t))$ for each $k \in \mathcal{Q}$. Then, it checks whether $r_3 \leq a_\mu(X^{c}(t)) / \overline{a_\mu}$ (respectively, $r_3 \leq a_\mu(X^{c + \epsilon_ke_k}(t)) / \overline{a_\mu})$ in order to update $X^{c}(t)$ (respectively, $X^{c + \epsilon_ke_k}(t)$). 

We emphasize that the selection of reaction firing in lines~\ref{begin-selection-loop} -~\ref{end-selection-loop} produces the exact marginal distribution for states $X^{c}(t)$ and $X^{c + \epsilon_ke_k}(t)$ for $k \in \mathcal{Q}$. A sketch of the correctness proof is as follows. We have for each selection step a candidate reaction $R_\mu$ is selected with probability $\overline{a}_\mu / \overline{a_0}$ and its waiting time is exponentially distributed with rate $\overline{a}_0$. Now consider the particular state $X^{c + \epsilon_ke_k}(t)$. The accepted probability of $R_\mu$ is $a_\mu(X^{c + \epsilon_ke_k}(t)) / \overline{a_\mu}$. The probability that $R_\mu$ is selected and accepted is thus $a_\mu(X^{c + \epsilon_ke_k}(t)) / \overline{a_0}$. Let $l_k$ be the number of trails until a candidate reaction is accepted to update the state $X^{c + \epsilon_ke_k}(t)$. The firing time at which the accepted candidate $R_\mu$ updates state $X^{c + \epsilon_ke_k}(t)$ is the sum of $l_k$ exponential random numbers, thus following the $\texttt{Erlang}(l_k, \overline{a}_0)$ distribution. From this point, the correctness argument of RSSA~\cite{TPZ_1} can be adapted to prove $R_\mu$ selected correctly with probability proportional to $a_\mu(X^{c + \epsilon_ke_k}(t))$, which ensures the exact marginal distribution of $X^{c + \epsilon_ke_k}(t)$.

The selection is repeated until there is a species whose population moves out of the fluctuation interval due to reaction firings. In this case, a new fluctuation interval as well as the propensity bounds of reactions should be updated. Line~\ref{begin-update} -~\ref{end-update} performs the update of propensity bounds when there existing species move out of their fluctuation interval. RFD keeps track of  species that should update their fluctuation intervals by the set $\textrm{UpdateSpeciesSet}$, which is initialized to be an empty set at the beginning in line~\ref{update-species-set}. For each species $S_i \in \textrm{UpdateSpeciesSet}$, a new fluctuation interval $[\underline{X}_i, \overline{X}_i]$ that bounds all populations of the species in all states is computed. Then, reactions affected by $S_i$, denoted by the set $\textrm{ReactionsAffectedBy}(S_i)$, is extracted from SR graph $\mathcal{G}$. For each $R_j \in \textrm{ReactionsAffectedBy}(S_i)$, its new lower bound $\underline{a_j}$ and upper bound $\overline{a_j}$ is recomputed. 

\section{Numerical examples}
\label{sec:4}
We report in this section the numerical results by our RFD algorithm in comparison with CRN and CFD algorithms. For the implementations of CRN and CFD, we use the dependency graph~\cite{GB_1} to decide which reactions should update their propensities when a reaction fires. All algorithms in this section are implemented in Java and run on a Intel i5-540M processor. The implementation of algorithms as well as the benchmark models are freely available at the url \url{http://www.cosbi.eu/research/prototypes/rssa}. We compare these methods in two models that are: the birth-death process and the Rho GTP-binding protein model. The former is a simple model, where the exact form of the sensitivity analysis is available, while the latter case is a large model where simulation must be used to perform sensitivity analysis. 

\subsection{Birth death process}
The birth-death process is a simple model, but is commonly found in applications. The model has two reactions that describe the producing and consuming of a species S. The reactions of the model are listed in Eq.~\ref{birth-death-process}.	
\begin{equation}
    \varnothing \stackrel{c_1}{\rightarrow} S \stackrel{c_2}{\rightarrow} \varnothing
\label{birth-death-process}		
\end{equation}
The species S is created with rate $c_1$ and degraded with rate $c_2$. The propensities of reactions in birth-death process is assumed to follow mass-action kinetics, hence $a_1 = c_1$ and $a_2 = c_2\#S$. For this model, we focus on sensitivities of the population of species S at time $t$ due to reaction rates. 

Let $s_0$ be the initial population of S at time $t = 0$. For this model, the population of species S at time $t$ can be computed analytically~\cite{Jahnke_2007} as the sum of Binomial distribution $Bin(s_0, p)$ and Poisson distribution $Po(\lambda)$ where $p = e^{-c_2t}$ and $\lambda = (c_1/c_2)(1 - e^{-c_2t})$. The expected value of number of species S at a time $t$ is thus given by
\begin{equation}
    \mathbb{E}[\#S(t)] = s_0e^{-c_2t} + (c_1/c_2)(1 - e^{-c_2t})
\end{equation}

\begin{figure}[!htb]
	\centering
	\includegraphics[scale=0.8]{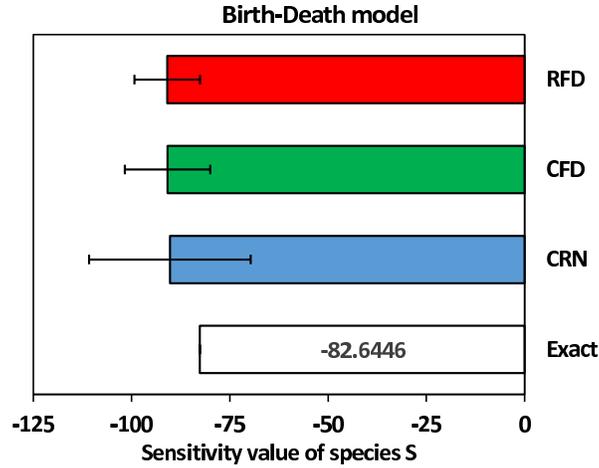}
	\caption{Sensitivities of the expected value of population of species S by CRN, CFD and RFD methods in comparison with exact value. The nominal reaction rate constant $c_2$ is increased by $\epsilon_2 = 10\%$. The sensitivity values are obtained by $1000$ runs of these methods each with simulation time $T_{max} = 100$.}
	\label{fig:sensitivity_1000_runs_birth-death}       
\end{figure}

\begin{figure}[!htb]
	\centering
	\includegraphics[scale=0.8]{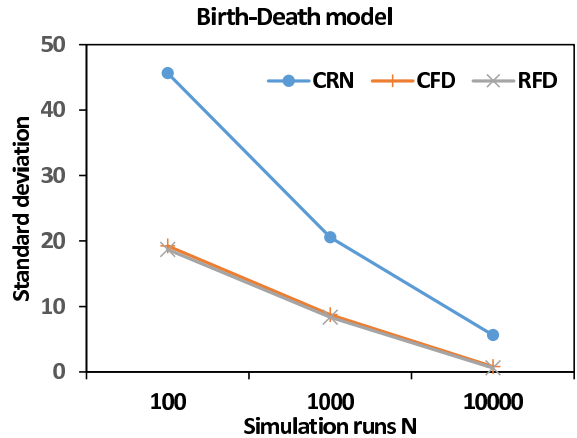}
	\caption{Standard deviations in estimating sensitivities of the expected value of population of species S by perturbing the reaction rate constant $c_2$ by an amount $\epsilon_2 = 10\%$ using CRN, CFD and RFD methods with different number of simulation runs.}
	\label{fig:std_estimator_birth-death}       
\end{figure}

For the computation of the sensitivities of the population of species S by CRN, CFD and RFD, the nominal rates of reactions are set to $c_1 = 100$ and $c_2 = 1$. The initial population of S is set $s_0 = 100$ and the simulation time is $T_{max} = 100$. First, we compute the sensitivities of the population of species S by increasing the reaction rate constant $c_2$ an amount of $\epsilon_2 = 10\%$ of this rate constant. We remark that the perturbation size in this section is defined as the percentage of change rather than the absolute value. The percentage is used in order to normalize the perturbation sizes~\cite{MOV_2012}. For the simulation of RFD, the fluctuation rate $10\%$ is applied to compute the fluctuation interval of species S. Figure~\ref{fig:sensitivity_1000_runs_birth-death} depicts the estimated sensitivity of the population of species S by CRN, CFD and RFD by $N = 1000$ simulation runs. Figure~\ref{fig:std_estimator_birth-death} gives the standard deviation for the estimators with varying the number of simulation runs. The figures show that the variances obtained by CFD and RFD are better than CRN. 

\begin{figure}[!htb]
	\centering
	\includegraphics[scale=0.8]{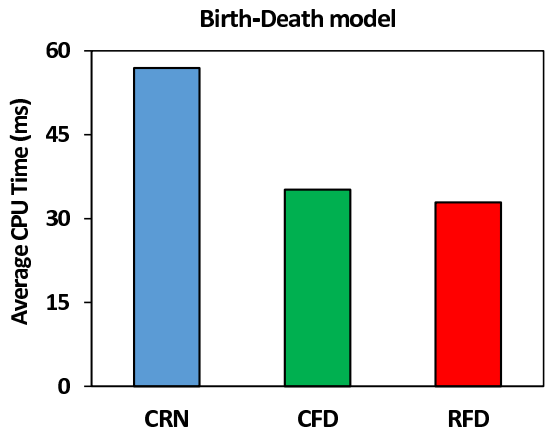}
	\caption{Performance of CRN, CFD and RFD methods in computing sensitivities of species S where the nominal reaction rate constant $c_2$ is increased by $\epsilon_2 = 10\%$.}
	\label{fig:performance_single_perturbation_birth-death}       
\end{figure}

Figure~\ref{fig:performance_single_perturbation_birth-death} shows performance of CRN, CFD and RFD in computing the sensitivities of the population of species S by increasing the reaction rate constant $c_2$ an amount of $\epsilon_2 = 10\%$. RFD has a similar performance as CFD, while it is about $2$ times faster than CRN. The performance gain by RFD in comparison with CRN is obtained by reducing the number of simulation steps and the number of propensity updates. The number of simulation steps, hence the number of propensity updates, performed by CRN is $4.0\times10^4$ and by CFD is $2.06 \times10^4$. RFD performs $2.21\times10^4$ simulation steps, but only has to update propensity bounds about $160$ times. 

Table~\ref{tab:sen-perf-rfd} shows the effect of the fluctuation rate, hence the values of propensity bounds, to the estimated sensitivity and performance of RFD. The results are obtained by $1000$ runs on the birth death process where reaction rate constant $c_2$ is increased by $\epsilon_2 = 10\%$. The table shows that the choice of fluctuation rate does not affect the sensitivity estimated by RFD, but only its performance. We note that the performance of RFD when the fluctuation rate is very small or very large is the worst. The small fluctuation rate, hence tight propensity bounds, causes many updates. The large fluctuation rate, hence loose propensity bounds, leads to many rejections. Both cases negatively affect the performance of RFD. In this experiment, the fluctuation rate $10\%$ gives the best performance.  

\begin{table}
\centering
\caption{Sensitivity and performance of RFD with varying fluctuation rate in simulating the birth death process where reaction rate constant $c_2$ is increased by $\epsilon_2 = 10\%$.}
\label{tab:sen-perf-rfd}       
\renewcommand{\arraystretch}{1.2}
\begin{tabular}{|r|l|r|}
          \hline  
			    Fluctuation rate  & Sensitivity of S & Average CPU Time (ms)\\
          \hline
				  $ 1\% $ & $-90.18 \pm 8.242$ & 48.39\\
				  $10\% $ & $-90.05 \pm 8.345$ & 34.65\\
				  $20\% $ & $-90.06 \pm 8.021$ & 37.76\\
				  $30\% $ & $-90.14 \pm 8.180$ & 37.99\\ 
				  $90\% $ & $-90.03 \pm 8.438$ & 42.12\\
					\hline             
\end{tabular}
\end{table}

\begin{figure}[!htb]
	\centering
	\includegraphics[scale=0.8]{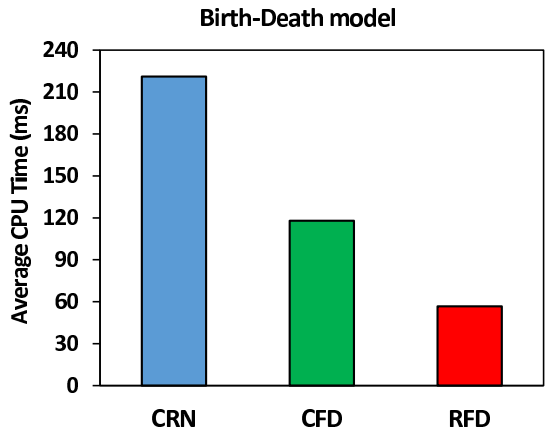}
	\caption{Performance of CRN, CFD and RFD methods in computing sensitivities of species S where both the reaction rate constant $c_1$ is increased by $\epsilon_1 = 1\%$ and the reaction rate constant $c_2$ is increased by $\epsilon_2 = 10\%$. The CRN and CFD methods have to repeat the computation $4$ times corresponding to $4$ combinations of perturbation parameters, while RFD performs simultaneously in one run.}
	\label{fig:performance_simultaneous_perturbation_birth-death}       
\end{figure}

For the second experiment, we repeat the sensitivity analysis of the population of species S by simultaneously perturbing both the reaction rate constant $c_1$ by $\epsilon_1 = 1\%$ and the reaction rate constant $c_2$ by $\epsilon_2 = 10\%$. For the simulation of RFD, the fluctuation rate $10\%$ is applied to compute the fluctuation interval of species S. The performance of algorihtms is averaged by $N = 1000$ simulation runs. In this experiment, because there are $4$ combinations of perturbation parameters (one of such combinations is shown in the previous experiment in Figs.~\ref{fig:sensitivity_1000_runs_birth-death} -~\ref{fig:performance_single_perturbation_birth-death}), CRN and CFD have to repeat the computation $4$ times corresponding to each combination. In contrast, RFD is able to perturb two reaction rates simultaneously. Figure~\ref{fig:performance_simultaneous_perturbation_birth-death} shows performance of CRN, CFD and RFD by simultaneously perturbing both the reaction rate $c_1$ and the reaction rate $c_2$. The figure shows that RFD is significantly more efficient than CRN and CFD. Specifically, the computational time of CRN and CFD is thus nearly $4$ times increased in comparison with the case where only one parameter is perturbed in the previous experiment. RFD in this setting performs $2.25\times10^4$ simulation steps which are similar to the case where only one parameter is perturbed. The increased computational time of RFD in comparison with the case where only one parameter is perturbed is due to more states keeping track. The result is the performance of RFD is about $3.89$ times and $2$ times faster than CRN and CFD, respectively. 

\subsection{Rho GTP-binding protein model}
We use the model of Rho GTP-binding proteins~\cite{KCCGP_2009,KL_2013} to demonstrate the computational efficiency of RFD in applying to large models. The Rho GTP-binding proteins constitute a subgroup of the Ras super-family of GTP hydrolases (GTPases) that regulate the transmission of external stimuli to effectors. The Rho GTP-binding protein cycle switches between inactive and active states depending upon binding of either GDP or GTP to the GTPases, respectively. The cycle is controlled by two regulatory proteins: guanine nucleotide exchange factors (GEFs) and GTPase-activating proteins (GAPs). GEFs promote the GDP dissociation and GTP binding, hence producing the activation of the GTPase. In contrast, GAPs stimulate the hydrolysis of the bound GTP molecules, hence transferring the GTPase back to the inactive state. In the active state, Rho GTP-binding proteins interact and activate downstream effectors.

Table~\ref{tab:Rho-GTP-binding-model} lists the reactions and the rates of the Rho GTP-binding protein model. In the model, R denotes the Rho GTP-binding protein in nucleotide free form and RD and RT denote its GDP and GTP bound forms, respectively. A and E denote GAP and GEF, respectively. The model has $23$ reactions. The initial populations for species are $\#R = 1000$, $\#E = 776$ and $\#A = 10$, while it is zero for all other species.  

\begin{table}
\centering
\caption{Rho GTP-binding model}
\label{tab:Rho-GTP-binding-model}       
\renewcommand{\arraystretch}{1.2}
\begin{tabular}{|l|l|}
          \hline  
			    Reaction                      & Rate        \\
          \hline
				  $R_1$: A + R  $\rightarrow$ RA 						& $c_1 = 1$ \\
				  $R_2$: A + RD $\rightarrow$ RDA 				  & $c_2 = 1$ \\
				  $R_3$: A + RT $\rightarrow$ RTA 				  & $c_3 = 1$ \\
				  $R_4$: E + R  $\rightarrow$ RE 					  & $c_4 = 0.43$ \\ 
				  $R_5$: E + RD $\rightarrow$ RDE 				  & $c_5 = 0.0054$ \\
				  $R_6$: E + RT $\rightarrow$ RTE 				  & $c_6 = 0.0075$ \\
				  $R_7$: R $\rightarrow$ RD 								& $c_7 = 1.65 $ \\
				  $R_8$: R $\rightarrow$ RT 								& $c_8 = 50$ \\
				  $R_9$: RA $\rightarrow$ A + R 						& $c_9 = 500$ \\
				  $R_{10}$: RD $\rightarrow$ R 						& $c_{10} = 0.02$ \\
				  $R_{11}$: RDA $\rightarrow$ A + RD 			& $c_{11} = 500$ \\
				  $R_{12}$: RDE $\rightarrow$ E + RD 			& $c_{12} = 0.136$ \\
				  $R_{13}$: RDE $\rightarrow$ RE 					& $c_{13} = 6.0$ \\
				  $R_{14}$: RE $\rightarrow$ E + R 				& $c_{14} = 1.074$ \\
				  $R_{15}$: RE $\rightarrow$ RDE 					& $c_{15} = 1.65$ \\
				  $R_{16}$: RE $\rightarrow$ RTE 					& $c_{16} = 50$ \\
				  $R_{17}$: RT $\rightarrow$ R 						& $c_{17} = 0.02$ \\
				  $R_{18}$: RT $\rightarrow$ RD 						& $c_{18} = 0.02$ \\
				  $R_{19}$: RTA $\rightarrow$ A + RT 			& $c_{19} = 3$ \\
				  $R_{20}$: RTA $\rightarrow$ RDA 					& $c_{20} = 2104$ \\
				  $R_{21}$: RTE $\rightarrow$ E + RT 			& $c_{21} = 76.8$ \\
				  $R_{22}$: RTE $\rightarrow$ RDE 					& $c_{22} = 0.02$ \\
				  $R_{23}$: RTE $\rightarrow$ RE 					& $c_{23} = 0.02$ \\
					\hline             
\end{tabular}
\end{table}

\begin{figure*}
	\centering
	\includegraphics[scale=0.8]{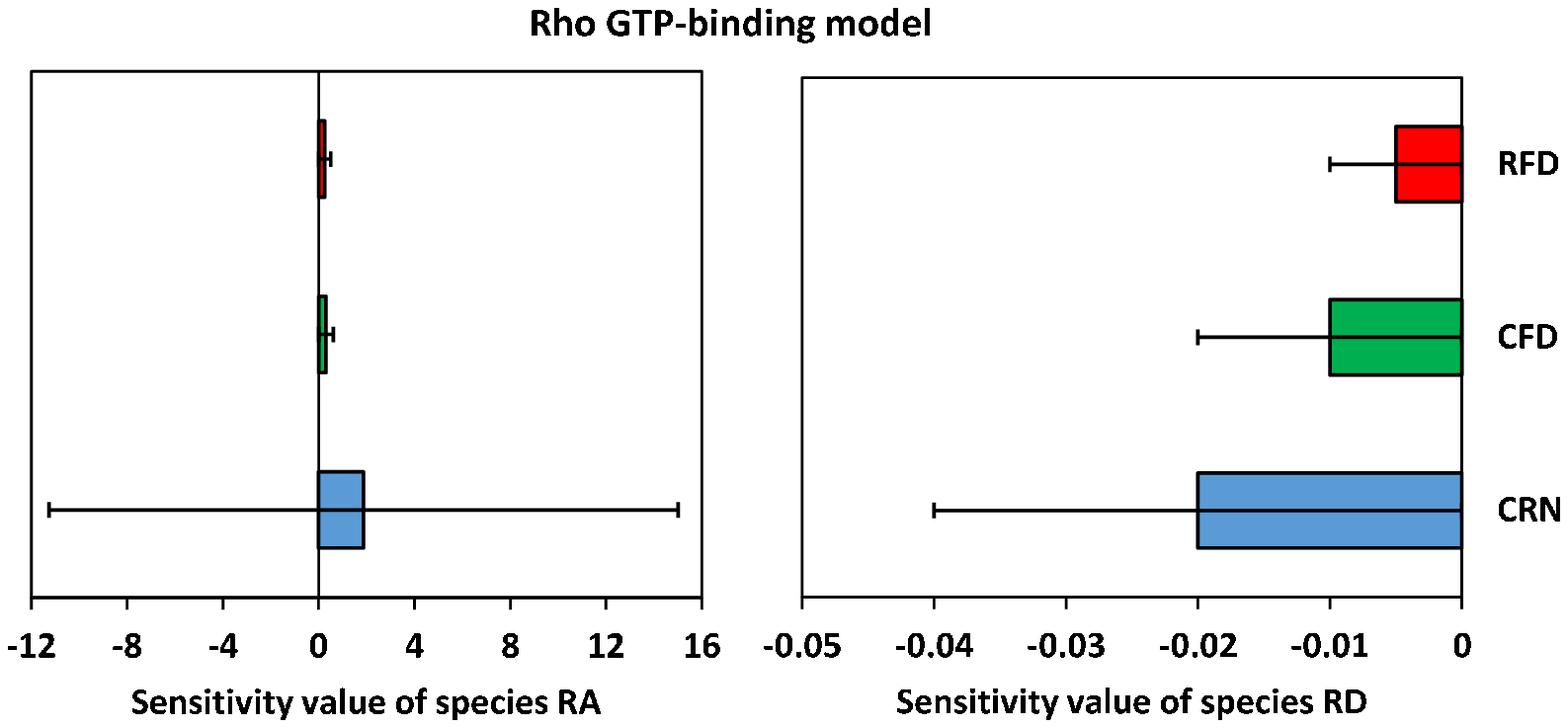}
	\caption{Sensitivities of the expected value of population of species RA and RD by CRN, CFD and RFD methods. The nominal reaction rate constant $c_1$ is increased by $\epsilon_1 = 20\%$. The sensitivity values are obtained by $1000$ runs of these methods each with simulation time $T_{max} = 10$.}
	\label{fig:sensitivity_1000_runs_rho-gtp-binding}       
\end{figure*}

\begin{figure}
	\centering
	\includegraphics[scale=0.8]{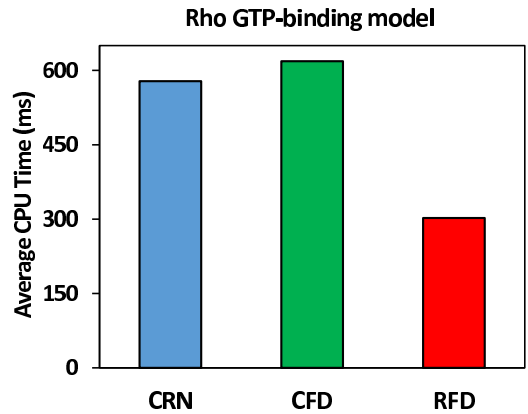}
	\caption{Performance of CRN, CFD and RFD methods in computing sensitivities of the expected value of population of species RA and RD where the reaction rate constant $c_1$ is increased by $\epsilon_1 = 20\%$.}
	\label{fig:performance_rho-gtp-binding}       
\end{figure}

Figure~\ref{fig:sensitivity_1000_runs_rho-gtp-binding} shows the sensitivities of the species RE and RD by increasing the rate $c_1$ by an amount $\epsilon_1 = 20\%$ and Figure~\ref{fig:performance_rho-gtp-binding} depicts the performance of CRN, CFD and RFD. The sensitivity values shown in Figure~\ref{fig:sensitivity_1000_runs_rho-gtp-binding} are obtained by performing $1000$ runs of algorithms with simulation time $T_{max} = 10$. These figures show that RFD is more efficient than both CFD and CRN in both the estimation of the sensitivities as well as performance. 

In Figure~\ref{fig:sensitivity_1000_runs_rho-gtp-binding}, the sensitivity values of species RE and RD estimated by CRN are less reliable due to the loose coupling of processes during the simulation although the reaction rate constant $c_1$ is less sensitive. The result is that the sensitivities estimated by CRN vary significantly. CFD and RFD helps to solve the problem efficiently by employing tightly coupling strategies. The nominal and perturbed processes by CFD and RFD jump together almost of the time during the simulation which result in a more reliable estimation by these algorithms. 

The performance plot in Figure~\ref{fig:performance_rho-gtp-binding} shows that RFD has the best performance, while CFD is the worst. The reason for the low performance of CFD in this experiment is due to the high computational cost for updating of propensities and related data structures of additional processes after reaction firings, even though CFD only performs $1.45\times10^5$ simulation steps which are a half as comparing with $2.88\times10^5$ steps by CRN and $2.53\times10^5$ steps by RFD, respectively. By reducing the propensity updates during the simulation, RFD significantly improves the performance. Specifically RFD only performs $2.13\times10^4$ propensity updates (about 9\% of its simulation steps). The simulation performance of RFD is $2.2$ and $2.7$ times faster than CRN and CFD, respectively.

\section{Conclusions}
\label{sec:5}
This paper proposed a new rejection-based finite difference (RFD) method for estimating sensitivities of biochemical reactions. Our method uses propensity bounds of reactions and the rejection-based mechanism to construct the estimator. By employing propensity bounds of reactions, the simulations of reactions with nominal and perturbed rates can be positively correlated, hence reducing of the variance of the estimator. The exactness of the simulation is preserved by applying the rejection-based mechanism. The advantage of using propensity bounds for coupling allows our method to simultaneously perturb many reaction rates at a time. The computational gain of our method is achieved by reducing the propensity updates during the simulation. Our rejection-based coupling is thus very promising for further investigation such as performing sensitivity analysis of reactions with time-dependent rates or computing high-order sensitivities used for optimization of biological processes. The weakness of our proposed rejection-based coupling is that the correlation may be loose if many simultaneous perturbations are applied and the dynamics of the nominal and perturbed processes significantly diverge. In such case, the propensity bounds will be large; thus, performance will suffer. In future work, we would investigate new strategies to mitigate the problem. It would also be interesting to understand whether the coupling is also affected.  
\end{document}